\documentclass[aps,prl,twocolumn,showpacs,groupedaddress]{revtex4}
\usepackage{amsmath}
\usepackage{graphicx}
\usepackage{dcolumn}
\usepackage{bm}

\begin{document}

\title{ Fine Details of the Nodal Electronic Excitations in 
Bi$_2$Sr$_2$CaCu$_2$O$_{8+\delta}$}  
\author{T. Valla}
 \email{valla@bnl.gov} 

\author{T. E. Kidd}
\altaffiliation {Physics Department, University of Northern Iowa, Cedar Falls, IA 50614-0150, USA}
\author{J. D. Rameau}
\author{H.-J. Noh}
\altaffiliation {School of Physics \& Center for Strongly Correlated Materials 
Research, Seoul National University, Seoul 151-742, Korea, and Department of Physics and Astronomy, 
Rutgers University, Piscataway, NJ 08854, USA}
\author{G. D. Gu}
\author{P. D. Johnson}
\affiliation{ Condensed Matter and Materials Science Department, Brookhaven 
National Laboratory, Upton, NY, 11973-5000}
\author{H.-B. Yang}
\author{H. Ding}
\affiliation{ Department of Physics, Boston College, Chestnut Hill, MA 02467}
\date{\today}

\begin{abstract}
Very high energy resolution photoemission experiments on high quality samples of 
optimally doped Bi$_2$Sr$_2$CaCu$_2$O$_{8+\delta}$ show new features in the 
low-energy electronic excitations. A marked change in the binding energy and 
temperature dependence of the near-nodal scattering rates is observed near the 
superconducting transition temperature, $T_C$. The temperature slope of the 
scattering rate measured at low energy shows a discontinuity at ~$T_C$. In the 
superconducting state, coherent excitations are found with the scattering rates 
showing a cubic dependence on frequency and temperature. The superconducting 
gap has a $d$-wave magnitude with negligible contribution from higher harmonics.  
Further, the bi-layer splitting has been found to be finite at the nodal point.
\end{abstract}


\maketitle
   High temperature superconductivity continues to present some of the biggest challenges in 
materials science today. What is the nature of the low energy excitations and what is the pairing 
mechanism that leads to the superconductivity? In attempting to answer these questions, 
angle-resolved photoemission spectroscopy (ARPES), with high resolution in both energy 
and momentum, has emerged as one of the leading techniques for the study of strongly correlated materials, 
including the high $T_C$ superconductors. Indeed, the demonstration that the $k-$dependence of the amplitude of the 
superconducting gap in these materials is consistent with $d-$wave symmetry represents an 
important contribution of the technique to the field \cite{Damascelli}. More recently, it has been 
shown that ARPES is an excellent probe of the single particle self-energy – a quantity that 
reflects the QP's interactions within the system and manifests itself in a renormalization 
of the single-particle dispersion and structure in the spectral width \cite{Moly}. The discovery of 
the mass renormalization or "kink" in the dispersion in Bi$_2$Sr$_2$CaCu$_2$O$_{8+\delta}$ 
has lead to renewed speculation about the origin of high 
temperature superconductivity (HTSC) and the possibility that the observed renormalization 
reflects coupling to some boson involved in the pairing \cite{Science}. Indeed, the "kink" has 
become one of the central issues in subsequent ARPES work, with considerable controversy 
regarding its source \cite{Damascelli, Kaminski, Lanzara, Johnson, Kim}. Is it related to 
the presence of spin excitations or does it reflect an interaction with phonons 
or indeed any other collective mode? Note that this is not an easy issue to 
resolve as the various energy scales are nearly identical. Our earlier studies of 
the doping and momentum dependence pointed to spin fluctuations \cite{Johnson}, while some 
papers favour phonons \cite{Damascelli, Lanzara} as the source of mass enhancement. 
   
A second important question that remains open is the detailed $k-$dependence of the 
superconducting gap $\Delta(\textbf{k})$. Although the momentum dependence 
of the gap amplitude was one of the first ARPES contributions to the field, 
\cite{Damascelli} the $k-$dependence from the near-nodal region has not previously 
been measured with sufficient precision. Early work showed that 
the contribution of higher $d-$wave harmonics grew as the doping was reduced \cite{Mesot}. Some recent papers suggest a significant 
contribution of higher harmonics in the cos$k_x$-cos$k_y$ distribution, even for samples relatively 
close to optimal doping \cite{Borisenko}. However, low-temperature thermal transport 
measurements suggest that the near nodal gap scales with the maximal gap $\Delta_0$ as measured 
in ARPES \cite{Damascelli} or tunnelling \cite{Renner}, even in severely underdoped 
samples \cite{Sutherland}. A detailed $k-$dependence of the gap is crucial as it determines the 
range of pairing interactions in real space. Further if the gap is a superposition of two 
competing orders, the $k-$dependence of the gap could provide insights into the 
relationship between them.
   
Finally, even though it is generally believed that the normal state of the cuprate 
superconductors is not a Fermi-liquid, the appearance of well defined quasiparticles (QPs) in the 
superconducting state has been well documented in other experimental techniques such as thermal 
conductivity \cite{Sutherland}, microwave \cite{Hosseini} and scanning tunnelling 
spectroscopy \cite{Hoffman}. However, there are conflicting ARPES results on the existence of 
nodal QPs \cite{Science, Kaminski2000}. In particular, one of our earlier 
studies reported that the spectral width in the nodal region was almost completely insensitive to the 
superconducting transition \cite{Science}. Theoretical studies indicate that in the superconducting 
state, phase space restrictions should 
result in a cubic dependence on $T$ and $\omega$ \cite{Dahm}.

In the present letter, we focus on these issues again and show that the sample quality and 
experimental resolution are crucial factors in resolving "fine details" of the electronic structure near 
the nodal point. We show that with sufficient resolution and high-quality samples, 
the ARPES experiments detect a marked change in the nature of the near-nodal excitation at the 
superconducting transition. Coherent states are found in the superconducting state, with the QP 
width being less than its energy. Below $T_C$ the scattering rate displays a cubic dependence on 
$T$ and $\omega$ up to $\sim75$ meV. In the normal state, the scattering rate varies 
as $\omega$ for $\omega> T$, and as $T$, for $T<\omega$. Further, we have found that the gap 
function near the node may be correlated with the amount of elastic scattering. In the highest 
quality samples the superconducting gap amplitude has a typical "V" profile, but with increasing elastic scattering the 
near-nodal gap becomes reduced and convex. We have also found a finite bi-layer splitting at the 
nodal point.

   The experiments were carried out on a Scienta SES-2002 electron spectrometer at beam line 
U13UB of the National Synchrotron Light Source. The combined instrumental energy resolution 
was set either to $\sim5$ meV or to 10 meV as indicated in the text. The angular resolution was 
better than $\pm0.003$\AA$^{-1}$ at the 16.7 eV photon energy used in the present study. Samples, 
grown by travelling solvent floating zone method, were mounted on a liquid He cryostat and 
cleaved \textit{in-situ} in the UHV chamber with a base pressure $6\times10^{-9}$ Pa. The 
temperature was measured using a calibrated silicon sensor mounted near the sample. The 
photoemission spectra were analysed using both energy distribution curves (EDC) and momentum 
distribution curves (MDC).
\begin{figure}
\begin{center}
\includegraphics [width=.9\columnwidth,angle=0]{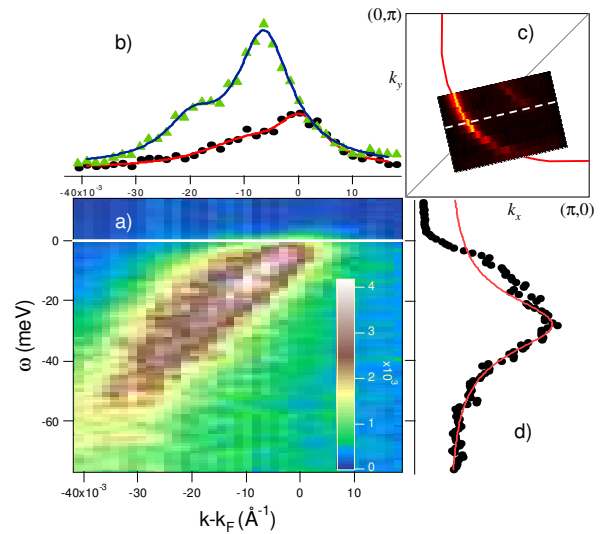}
\end{center}
\caption{\label{fig1} a) High-resolution photoemission spectrum of optimally doped Bi$_2$Sr$_2$CaCu$_2$O$_{8+\delta}$ at 8 K, 
 along the dashed line in the Brillouin zone, as indicated in (c). 
b) MDCs at $\omega=0$ (circles) and at $\omega=-10$ meV (triangles).
c) Fermi surface reconstructed from a series of spectra along momentum lines parallel to the dashed line. 
d) EDC at a fixed momentum ($k-k_F=0.015$\AA$^{-1}$). 
The solid line shows a Lorentzian fit to the near-peak and the high-energy side of the bonding state peak.}
\end{figure}

   Fig. 1 shows the photoemission spectrum recorded with 5meV resolution in the 
superconducting state along the momentum line as indicated in the inset. Several lines were 
scanned in momentum space parallel to this line ($30^\circ$ relative to $\Gamma$Y) allowing the 
Fermi Surface in the near-nodal region to be reconstructed. A bi-layer splitting that is finite and 
measurable is obvious in both the MDCs and EDCs. At the nodal point the splitting is visible only with the energy resolution of 
10 meV or better. Measured with 5 meV resolution, we determine by fitting with two 
Lorentzian peaks that the width of the bonding component is $\Delta k_B\approx 0.009$\AA$^{-1}$ and 
the splitting between the two components is $k_B-k_A\approx 0.008$\AA$^{-1}$ 
at the nodal point. These values represent the components perpendicular to the Fermi surface. 
Similar values for the widths and splitting have been reported elsewhere \cite{Kordyuk}. 

In Fig. 2(a) we show the $k$-dependences of momentum width of the bonding state at the Fermi 
surface and of the momentum splitting between the bonding and anti-bonding states. Again, the components perpendicular to the Fermi surface are shown, this time as 
measured with 10 meV resolution. As anticipated, the splitting between the states becomes larger as 
one moves away from the nodal point. As an indication of the quality of the present sample, \textbf{A}, we also show the near 
nodal width from another optimally doped sample, \textbf{B}, measured at the same temperature 
and with same experimental resolution. The larger widths reflect a higher level of 
impurity or defect scattering. In Fig. 2(b) we show the amplitudes of the 
superconducting gaps, $\Delta(\textbf{k})$, as measured using the leading edge of a spectrum integrated over a 
finite momentum range, typically $\pm 0.05 $\AA$^{-1}$ around $k_F$ along the measured momentum 
lines. We have also used different methods of extracting the gap amplitude \cite{Gap}, resulting 
in the same $k-$dependence. The gap function of the sample with sharp nodal states (\textbf{A}) 
clearly has a V-like profile, while the sample with broader states (\textbf{B}) shows a convex gap 
function (flatter gap) near the node. The function $\Delta_0[B$cos$(2\phi)+(1-B)$cos$(2\phi)]$ 
has been used to describe the anisotropy of the gap with $\Delta_0$ the maximum gap amplitude 
\cite{Mesot}. The V-shape is characteristic of a pure cos$(2\phi)$ with little or no contribution 
from higher harmonics and points to the dominance of nearest neighbor interactions in the pairing 
in sample \textbf{A}. However, with an increasing contribution from impurity scattering, 
the gap amplitude near the node gets reduced, states become broader, and any fine detail of the type 
discussed below becomes unresolved. 
\begin{figure}
\begin{center}
\includegraphics {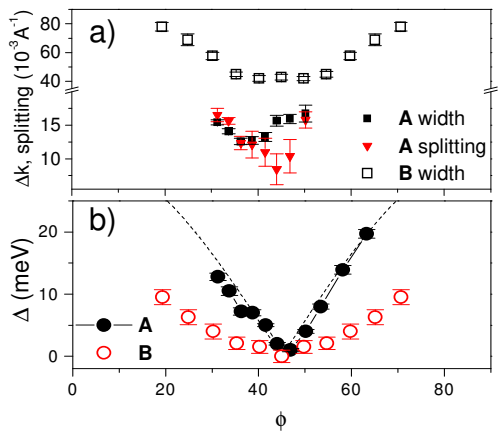} %[width=.75\columnwidth,angle=0]{fig.eps}
\end{center}
\caption{\label{fig2} 
a) Momentum width of the Fermi surface as a function of angle $\phi$ 
for two optimally doped samples (\textbf{A} and \textbf{B}). The momentum splitting between bonding 
and anti-bonding states for the sample with sharper states (\textbf{A})
 is also plotted. Components normal to the Fermi surface are shown. 
b) Superconducting gap for samples \textbf{A}, \textbf{B}.}
\end{figure}   
   
Fig. 3(a) shows the effects of the superconducting transition on the scattering rates for 
sample \textbf{A}. A normal state spectrum was recorded with 10meV resolution and fitted with a single 
Lorentzian lineshape. Two spectra in the superconducting state were recorded consecutively with 
10 meV and 5 meV resolution and fitted with a single Lorentzian peak and with two Lorenztian 
peaks, respectively. It is clear that most of the change between the normal and superconducting 
states occurs in the low binding energy range, below $\sim75$meV. If taken with high resolution, 
the superconducting spectrum shows a smooth variation of the MDC width in that region, with no 
indication of any prominent structure up to $\sim75$meV, where the functional form changes into a 
nearly linear dependence. In the latter regime, the scattering rate is almost unaffected by the 
superconducting transition \cite{ImS}. The low energy region recorded with high resolution 
may be fitted with an $\omega^3$ dependence (Fig. 3(b)), characteristic of electron-electron scattering in the presence 
of a $d-$wave gap \cite{Dahm}. It is clear that 
the fit does not catch the lower frequencies accurately. In this region a linear dependence seems more appropriate. 
The total scattering rate or inverse lifetime reflects both elastic and 
inelastic contributions, the former from impurities, and the latter from electronic and bosonic 
scattering processes. As such, the finite scattering rate at the lowest temperatures at $\omega=0$ reflects the elastic scattering from 
impurities or defects. Furthermore it has been suggested elsewhere \cite{Dahm, Zhu} that the 
lowest frequency region is dominated by the quasi-linear dependence of weak elastic scattering 
from out of plane impurities, consistent with our observations.
\begin{figure}
\begin{center}
\includegraphics {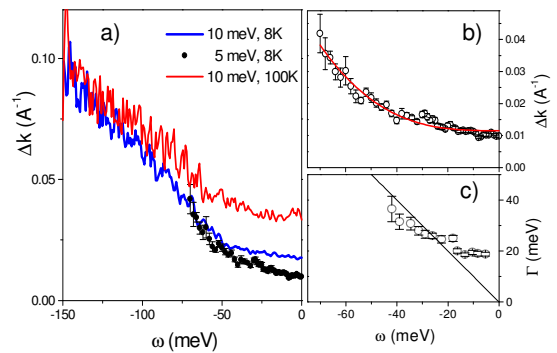} 
\end{center}
\caption{\label{fig3} 
a) Width of the MDC peak in the nodal region in the normal and superconductiong state recorded
with 5 meV and 10 meV energy resolution, as indicated.
b) The 5meV resolution data from (a), fitted with a cubic fit (solid line). 
c) EDC width as a function of binding energy at 8K. The straight line indicates the "QP" boundary, $\Gamma(\omega)=|\omega|$.}
\end{figure}
   
We also show in Fig. 3(c) the width of the lower (bonding) state as determined from EDCs 
recorded with 5 meV resolution and fitted with Lorentzian peaks. Fitting the EDCs with 
Lorentzians works well because Im$\Sigma$ is a slowly varying 
function, and its value is small at low energies $(|\omega|\le50$meV). As indicated in the figure, it 
is clear that over a finite range of energies near the Fermi level, the width of the state 
is less than its energy, in agreement with recent laser induced ARPES 
studies \cite{Koralek}, an indication that nodal QPs exist in the superconducting state. 
However, this clear evidence of QPs appears to be true only at the low 
temperatures. In the vicinity of $T_C$, the criterion that the 
width of the state is less than its energy is satisfied only marginally and over a very 
narrow range of energies even after subtracting the contribution from impurity 
scattering and energy resolution. This is markedly different from the observations on metallic 
systems such as our study of molybdenum \cite{Moly}.

More detailed temperature dependent studies of the electron self-energies for the same 
momentum line (from Fig. 1) are shown in Fig. 4. In 4(a) the MDC deduced dispersions are 
shown for several temperatures, while 4(b) shows the corresponding energy dependence of the 
momentum widths. The $T$-dependence of the width at the Fermi level is shown in 4(c). 
For this analysis all the spectra were taken with 10 meV total resolution and the MDCs were 
fitted with a single Lorentzian. As such the widths at low $T$ are larger than the equivalent width 
in Fig. 3(b). Fig. 3(a) and figures 4(b) and 4(c) allow us to compare the scattering rates more 
directly above and below $T_C$. It is clear that the opening of the $d-$wave superconducting gap 
modifies the functional form of the nodal scattering rate, leading to the cubic dependence on 
binding energy at low frequencies as discussed above. Also, the MDC width at the Fermi level 
$(\omega=0)$ as a function of $T$ shows a discontinuity in slope at, or slightly above $T_C$. The 
discontinuity and a rapid narrowing below a certain temperature suggest the formation of coherent 
quasi-particles in the low-$T$ state, consistent with our observations discussed earlier. Below $T_C$ 
the spectral points in Fig. 4 (c) can be fitted with a cubic dependence as 
predicted for QPs scattering off spin fluctuations in a superconducting state with $d-
$wave gap \cite{Dahm, Zhu}.
\begin{figure}
\begin{center}
\includegraphics [width=.99\columnwidth,angle=0]{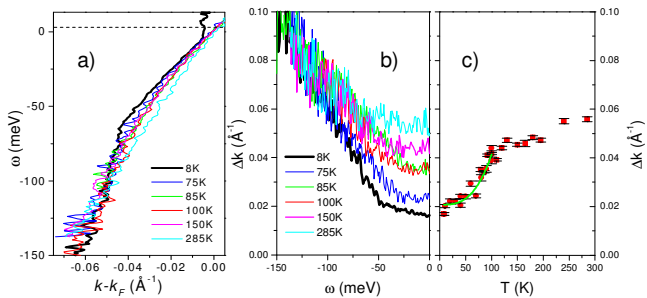}
\end{center}
\caption{\label{fig4} 
a) The MDC-derived dispersion along the line indicated in Fig. 1c at different temperatures. 
b) The corresponding MDC peak-widths. 
c) The $T$-dependence of MDC width at the Fermi level. The solid line is a cubic fit to 
the data below $T_C$.}
\end{figure}
      
Elsewhere it has been suggested that the kink or mass renormalization observed in the nodal 
direction in the cuprates reflects coupling to some optical phonon mode \cite{Damascelli, 
Lanzara}. Let us consider this suggestion. In the usual Migdal-Eliashberg approximation, the 
scattering rate or imaginary component of the self-energy is given by 
\begin{equation}
\mathrm{Im}\Sigma(\omega)\propto \int_{0}^{\omega}d\nu \frac{N(\nu)}{N_0}\alpha^2F(\omega -\nu)
\label{eq1}
\end{equation}  
where in the superconducting state with the order parameter having $d-$wave symmetry, the 
density of states (DOS) may be approximated by $N(\omega)/N_0\approx\omega/\Delta_0$ at low 
frequencies. Here $\Delta_0$ again represents the maximal superconducting gap energy and $N_0$ 
represents the normal state DOS. Examination of equation (1) shows that if the phonon mode 
energy (corresponding to the kink energy) is greater than $\Delta_0$, the scattering rate as a 
function of binding energy at energies below the mode will be insensitive to the superconducting 
transition. This is clearly inconsistent with figures 3(a) and 4(a,b,c) and \textit{essentially rules out the 
possibility that the kink reflects coupling to such a single Einstein mode}. The linear term and a 
finite zero-energy offset indicate, as already noted, that the lowest energy range is still dominated 
by impurity scattering. In that case, the low energy scattering rate would be affected by the 
transition. However, the change in the elastic scattering would be limited to energies inside the 
maximal superconducting gap $\Delta_0$, as it would reflect the redistribution of density of states 
upon transition. Here, the observation that there are changes at frequencies below the mode, but 
outside the gap energy $\Delta_0$, rules out the scattering from a single phonon mode. One simple 
explanation would be that the bosonic spectrum itself changes at transition. A redistribution of 
bosonic states at low energies, while conserving the total density (e.g. by opening of a gap at low 
energies) would strongly affect the electronic scattering rates at low energies, with minimal impact 
on those at high energies. We note in passing that any changes in the scattering rate or Im$\Sigma$ 
for the nodal excitations must be accompanied by equivalent changes in Re$\Sigma$ through 
causality. This is clearly visible in Fig 4(a) where the changes in Re$\Sigma$ affect the measured 
dispersions, in accordance with our previous study \cite{Johnson}. 
	
In summary we have demonstrated a marked change in both the energy and 
temperature dependence of the scattering rates for the low energy excitations in the nodal region on 
entering the superconducting state. As we have reported before \cite{Science} and as found in 
optical conductivity studies \cite{Tu} the states at higher binding energies show the same linear 
dependence on binding energy independent of whether the system is in the normal or 
superconducting state. This observation points to strong scattering in the electron-electron channel 
that reflects the entire band. The states at low frequencies have lifetimes that suggest they 
represent coherent Fermionic excitations. Further they appear sensitive to the presence of the 
superconducting gap and indeed the scattering rates have the anticipated cubic dependence on 
binding energy and temperature that reflect the restricted phase space in the $d-$wave 
gap. In addition, the observation that the scattering rate as a function 
of binding energy is modified on entering the superconducting state also provides evidence that the 
interaction responsible for this is electronic in origin. A high frequency phonon at 
$\omega_0$ would not affect the relaxation rate until $T$ or $\omega$ exceeds $\omega_0$.

We acknowledge useful discussions with Andrey Chubukov, Steve Kivelson, Maurice Rice, Doug 
Scalapino, John Tranquada, Alexei Tsvelik and Ziqiang Wang. The research work described in this 
paper was supported by the Department of Energy under Contract No. DE-AC02-98CH10886, and 
NSF DMR-0353108.


\begin{thebibliography}{99}
\bibitem{Gap} Gap amplitudes were also extracted from the peak (leading edge) position of EDCs 
at $k=k_F$, resulting in a small positive (negative) offset relative to those in Fig. 2b).
\bibitem{ImS} In the linear range, the imaginary component of the self energy, Im$\Sigma=(\Delta 
kv_0)/2$, varies as $\sim 0.79\omega$, if the bare velocity is set to $\sim3$eV/\AA.
\end{thebibliography}
\end{document}